\makeatletter
\@ifundefined{@parse@version@dash}{%
\def\@parse@version#1{\@parse@version@0#1}
\def\@parse@version@#1/#2/#3#4#5\@nil{%
\@parse@version@dash#1-#2-#3#4\@nil}
\def\@parse@version@dash#1-#2-#3#4#5\@nil{%
  \if\relax#2\relax\else#1\fi#2#3#4 }
}{}
\makeatother

\documentclass[%
 preprint,
 titlepage,
 superscriptaddress,
 amsmath,
 amssymb,
 aps,
 pra,
 showkeys,
 showpacs,
]{revtex4-2}
\usepackage{graphicx}
\usepackage{dcolumn}
\usepackage{bm}
\usepackage{hyperref}

\begin{document}
\title{Structure Property in Cu crystallization}

\author{Bobin Li}
\thanks{Corresponding author:gslibobin@lut.edu.cn}%
\affiliation{Department of Physics, Lanzhou University of Technology, Lanzhou 730050, China}
\date{\today}
\begin{abstract}
Phase transition is a central topic in condensed matter physics, all the time. In this paper, as a general representative of phase transition, the Cu crystallization is discussed. And some physical quantities is defined to quantificationally describe the structure property in Cu crystallization, such as diffusion property and symmetry so on. As a result, it is indicated that there are some interesting changes of structure property in Cu crystallization.
\end{abstract}

\maketitle

\section{Introduction}

With the progress of condensed matter physics, the phase transition becomes more and more important, because that it is common forms of being from a phase(configuration state) to another phase. In general, phase transition means a systemic change of condensed matter states of various particles. The classic phase transition is described by thermodynamic parameters, such as temperature and pressure so on. The system of phase transition follows the minimum condition of free energy in macroscopic states, and the microcosmic structure recombines again. And there is a discontinuously change for systemic macroscopic thermodynamic quantities or their derivatives.

When the liquids are cooled down, they will coagulate into solids, and, among them, the some will crystallize. The crystallization is a first order thermodynamic phase transition, and it will be affected by various factors. For example, in some cases, water does not form crystal until temperature below -40.0 celsius \cite{1,2}. And, under the high pressure, it will crystallize until temperature below -70.0 celsius \cite{3}. And there are some other peculiar phenomenons in crystallization. For instance, the most of crystallization is exothermic, meaning that pressure and energy are released when the liquid becomes a crystal. But Helium is a only known matter that does not exothermic when in crystallization\cite{4}, which indicates that the heat needs to be provided to make helium crystallize under some specific pressures\cite{5}. As the some time, there are some extremely crucial properties in crystallization that is symmetry. And the crystallization breaks the symmetry and affect the solid pattern of system \cite{6}. Before temperature of a liquid phase drops close to the freezing point, the system has spatial symmetry. But when temperature drops below the freezing point, the symmetry will be broken\cite{7,8} and the crystal structure will be determined. So symmetry breaking plays an important role in crystallization.

In this work, the molecular dynamics(MD) simulation is applied for Cu system, and then the particle trajectories, positions and velocity distribution is directly obtained by MD. According to the definition of peoperty parameters, we will get their patterns and then analyses them in Cu crystallization. Finally, it is obtained for macroscopic thermodynamic information of Cu system by statistical methods. As a conclusion, it is revealed that structure property is a interesting change in Cu crystallization.

\section{Property Parameters}
\subsection{The Radial distribution Function(RDF)}
In condense matter physics, the RDF describes the probability of an atom appearing nearby the given atom. It is a relatively efficient method to describe disordered structure. Generally, in the exploration of a system structure, it could be estimated for coordination number by RDF. And its result also is compared with the result of the X-ray diffraction by MD simulation \cite{9}. And definition as follows
\begin{equation}
g(r)=\frac{\overline{dn(r)}}{4\pi r^2 \rho}
\end{equation}
According to the $g(r)$ definition, when $r\rightarrow \infty$, the $g(r)\rightarrow 1.0$ in the disorder structure such as liquid. But in the order structure, when $r\rightarrow \infty$, there are also some peaks for anisotropism of it, and the density of peaks could partly reflect symmetry of system.

\subsection{Mean square displacement(MSD) and Diffusion coefficient}
In material science, the MSD could describe dynamic properties of continuous systems, and diffusion coefficient could reflects macroscopic diffusion features. And definition as follows
\begin{equation}
 MSD(t)=\frac{1}{N_m}\langle \sum^{N_m}_{j=1}[r_j(t)-r_j(0)]^2\rangle
\end{equation}

The systemic diffusion coefficient can be obtained. Definition as follows
\begin{equation}
 D=\lim_{t \rightarrow\infty} \frac{1}{6N_m t}\langle \sum^{N_m}_{j=1}[r_j(t)-r_j(0)]^2\rangle=\lim_{t\rightarrow\infty} \frac{1}{6 t}MSD(t)
\end{equation}
For limited system, the simulation time cannot be infinite. So we can only get D value with a limited evolution time, and then get diffusion coefficients as $t \rightarrow \infty$ by extrapolation.

The MSD and D both reflect the macroscopic diffusion characteristics of system structures.

\subsection{Bond orientational order parameter}
\begin{equation*}
 Q_l=[\frac{4\pi}{2l+1} \sum^{l}_{m=-l}|\bar{Q}_{lm}|^2]^{1/2}
\end{equation*}
The $Q_l$ reflects symmetry of a system. And the $Q_6$ is the most sensitive parameter to the fcc structure than other parameters(the crystal of Cu system is fcc). So we could analyse the symmetry of systemic structure by $Q_6$\cite{10}. And definition as follows
\begin{equation}
 Q_6=[\frac{4\pi}{13} \sum^{6}_{m=-6}|\bar{Q}_{6m}|^2]^{1/2}
\end{equation}

In this paper, we analyse all above property parameters in the process of change between liquid to solid state in Cu system. First, we build the fcc crystal structure with a Cu system of 500.0 atoms, and then make it melt to liquid state at 3000.0K. And the melting point is between 2000.0K to 2200.0K(It is not a perfect crystal).


\section{Results and Discussion}
\subsection{The Radial distribution Function(RDF) in Cu crystallization}
According to the data of Cu system, the $g(r)$ distribution is analysed in the NVT ensemble. The range of temperature is from 500.0K to 2500.0K, and the $g(r)$ distribution is shown in Cu system as follows.
\begin{figure}
\centerline{\includegraphics[width=13.0cm]{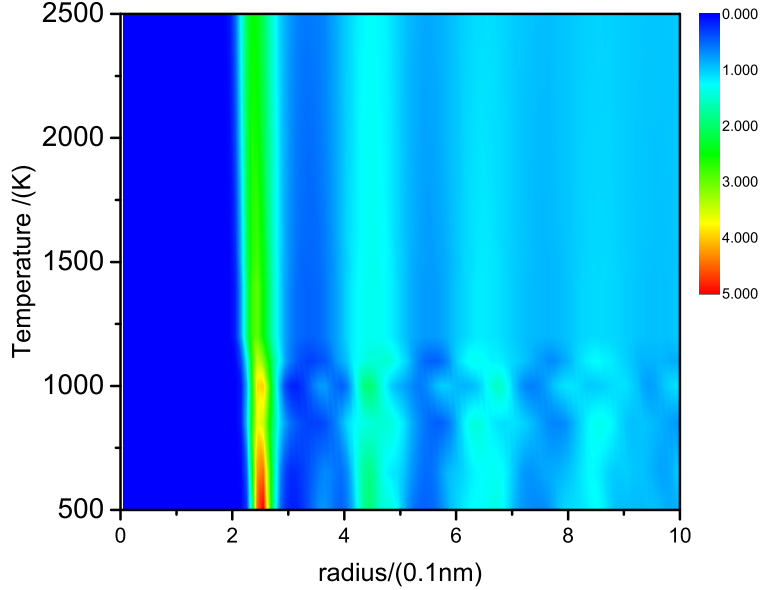}}
\caption{Under different temperature, the $g(r)$ distribution of Cu system.}
\end{figure}

When the temperature is between 2000.0K to 2500.0K, and, with a long distance r, the value of $g(r)$ tends to 1.0, which reflects systemic feature of disorder, and system is still in the state of liquid phase. With the decrease of temperature, the change of $g(r)$ become more and more sharp, and there appears more and more peaks in long distance. And it means that the systemic structure shifts from disorder phase to the ordered phase under low temperature. Meanwhile, the density of Cu system becomes more and more dense. And it indicated that the long-range order phase emerges in the process of crystallization.

\subsection{The Mean square displacement(MSD) and Diffusion coefficient of Cu syetem }
The MSD(t) describes the mean azimuth shift of Cu system with time, there is MSD as follows\cite{11}.
\begin{figure}
\centerline{\includegraphics[width=13.0cm]{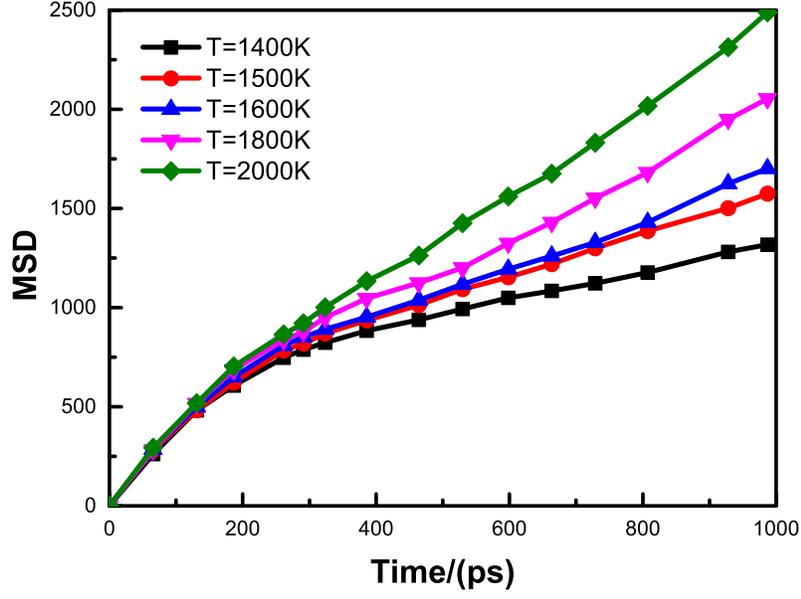}}
\caption{Under different temperature, the MSD(t) of Cu system}
\end{figure}
From FIG.2, when $t<100.0ps$, there is not collision between atoms, called ballistic inside area, so MSD(t) is proportional to mean square time. And when $200.0ps<t<400.0ps$, MSD does not change with time, due to cage effect. When $t>400.0ps$, MSD has linear relation with time.

The diffusion coefficient D(t) changes with temperature, and there is the D(t) as follows\cite{11}.
\begin{figure}
\centerline{\includegraphics[width=13.0cm]{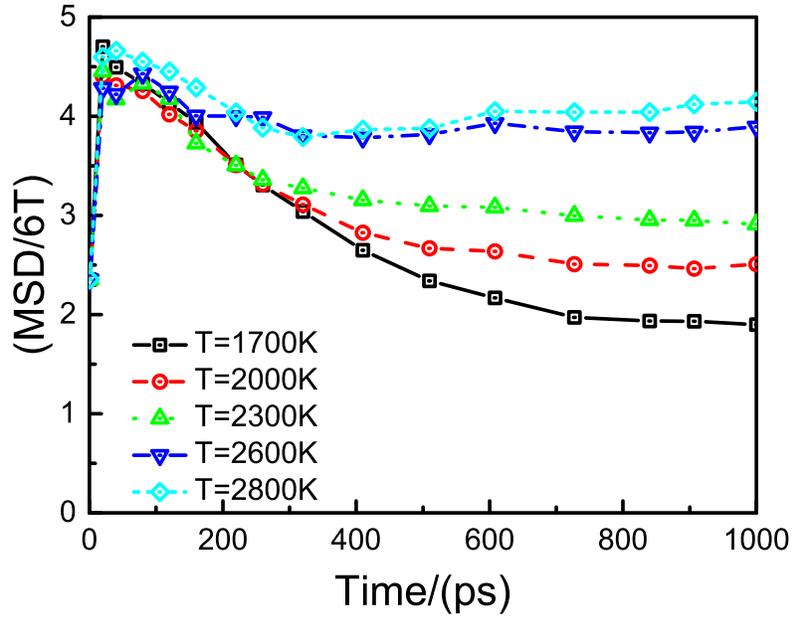}}
\caption{Under different temperature, the D(t) of Cu system. The value of $D(t \rightarrow \infty)$ is
$D(T=1700K)=1.92$,$D(T=2000K)=2.50$,$D(T=2300K)=2.93$,$D(T=2600K)=3.87$ and $D(T=2800K)=4.11$,the unit of $D(T)$ is ${nm}^2/(100ps)$}
\end{figure}
Diffusion coefficient D(t) is shown in FIG.3. And D value is got by extrapolation ($t \rightarrow \infty$), so $D(T=1700K)=1.92$,$D(T=2000K)=2.50$,$D(T=2300K)=2.93$,$D(T=2600K)=3.87$ and $D(T=2800K)=4.11$(${nm}^2/(100ps)$).
As a conclusion, the Diffusion coefficient D increases but is not more and more sensitive with the increase of temperature.

\subsection{The bond orientational order parameter $Q_6$ in Cu crystallization}
Generally, $Q_6$ of liquid is about 0.29, and perfect fcc crystal $Q_6=0.57$. When $Q_6<0.1$, it indicates that system is in a totally disorder state\cite{12}. And common crystal of Cu is standard fcc.
\begin{figure}
\centerline{\includegraphics[width=13.0cm]{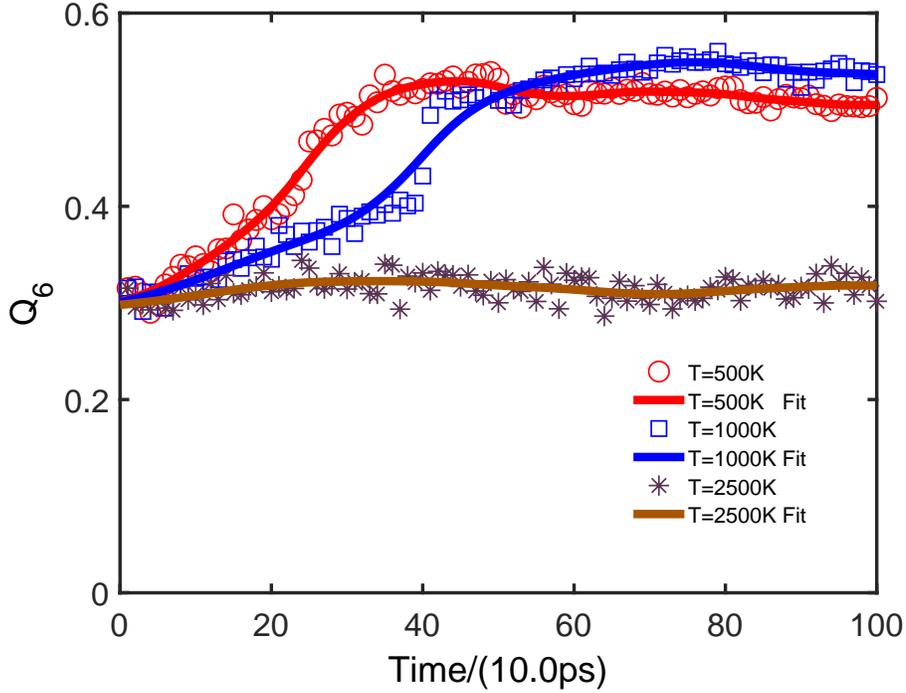}}
\caption{Under different temperature, the global $Q_6$ in Cu crystallization}
\end{figure}
When temperature is 2500.0K, there is no phase transition, and Cu system keeps to liquid phase. The system symmetry is invariant. When temperature is 1000.0K and 500.0K, the phase transition happens. Showing in FIG.4, system symmetry has a marked increase in the process of Cu crystallization. That is, the phase of Cu atoms tend to be orderly, the same as $g(r)$ showing. And there is a key, when temperature is 1000.0K, system has a longer time to finish crystallization than 500.0K, and it is obvious. But, after crystallization, the system symmetry of 1000.0K has a larger state than 500.0K. The reason is that the disorder is locked in solid state of Cu system, duo to cooling time is too short. And this case is like glass, there are many defects in Cu solid. Further, a detail is analysed in FIG.5.

\begin{figure}
\centerline{\includegraphics[width=13.0cm]{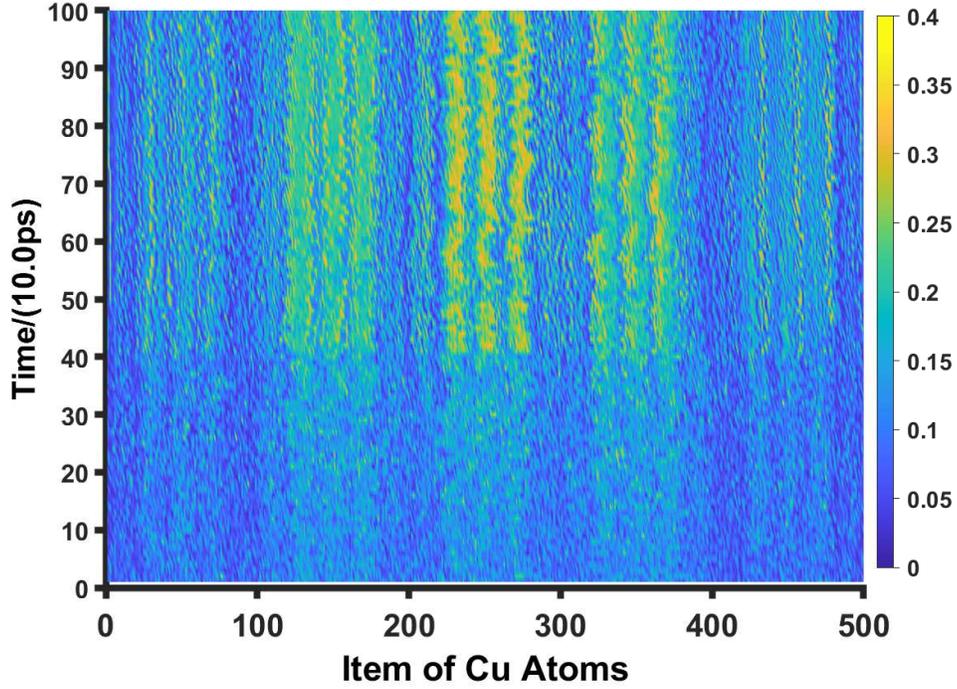}}
\caption{When temperature is 1000K, the distribution of local $Q_6$ in Cu crystallization}
\end{figure}

It describes the change of local symmetry with time in the process of crystallization. Before crystallization, basically, the local symmetry of Cu system is well-distributed in space. But, along with the growth of time, around $400.0ps$, the phase transition becomes obvious. And local symmetry is not uniform in space and time. Moreover, after crystallization, local symmetry is locked. And it is easy to understand, after crystallization, Cu atoms are fixed in a space point by potential, and do not migrate but only vibrate.

\section{Conclusion}
Firstly, we introduce some property parameters, including radial distribution function, mean square displacement, diffusion coefficient and bond orientational order parameter, to describe the phase transition of Cu system. By the method of molecular dynamics, it is shown that the correlation of systemic radial density changes from long-range to short-range and the diffusion capacity increases but is not sensitive, with increase of temperature,  in crystallization.

And the symmetry has a obvious change, whether system symmetry or local symmetry. In crystallization, average symmetry of Cu system increases, but it is affected by temperature. Before the crystallization, local symmetry basically is uniform in space-time, but, after crystallization it, is no longer uniform. If we want to get perfect Cu crystal, it is not enough to Cool down.

Finally, we hope that it is got for the information of material structure by neutron scattering experiments or X-ray diffraction. And combining molecular dynamics and experiments both is to explore new properties in materials.

\newpage
\bibliography{apssamp}

\end{document}